\documentclass[useAMS,usenatbib]{mn2e}
\usepackage{amsmath,amssymb}
\usepackage{graphicx}

\title[Spectroscopic Monitoring of A0535+26: I]{High-dispersion spectroscopic monitoring of the Be/X-ray binary A0535+26/V725 Tau I: The long-term profile variability}
\author[Y. Moritani et al.]{Y. Moritani$^{1}$\thanks{E-mail: moritani@kusastro.kyoto-u.ac.jp}, D. Nogami$^{2}$, A. T. Okazaki$^{3}$, A. Imada$^{4}$, E. Kambe$^{4}$, S. Honda$^{5}$, 
\newauthor
O. Hashimoto$^{5}$, Y. Ishino$^{1}$, Y. Suzuki$^{1}$ and J. Tanaka$^{1}$ \\
$^{1}$Department of Astronomy, Kyoto University,
 Sakyo-ku,  Kyoto 606-8502, Japan \\
$^{2}$Kwasan Observatory, Kyoto University,
 Yamashina-ku, Kyoto 607-8471, Japan \\
$^{3}$Faculty of Engineering, Hokkai-Gakuen University,
 Toyohira-ku, Sapporo 062-8605, Japan \\
$^{4}$Okayama Astrophysical Observatory, National Astronomical Observatory of Japan,
 3037-5 Honjo, Asakuchi, Okayama 719-0232, Japan \\
$^{5}$Gunma Astronomical Observatory,
 Takayama-mura, Gunma 377-0702, Japan 
} 
\begin{document}

\date{Accepted xxxx Xxx xx. Received xxxx Xxx xxx}

\pagerange{\pageref{firstpage}--\pageref{lastpage}} \pubyear{xxxx}

\maketitle

\label{firstpage}

\begin{abstract}
We report on optical high-dispersion spectroscopic monitoring observations of the Be/X-ray binary A0535+26/V725 Tau, carried out from November 2005 to March 2009.
The main aim of these monitoring observations is to study spectral variabilities in the Be disc, on both the short (a week or so) and long (more than hundreds of days) timescales, by taking long-term frequent observations.
Our four-year spectroscopic observations  indicate that the V/R ratio, i.e., the relative intensity of the violet (V) peak to the red (R) one, of the double-peaked H$\alpha$ line profile varies with a period of 500 days.
The H$\beta$ line profile also varies in phase with the H$\alpha$ profile.
With these observations covering two full cycles of the V/R variability, we reconstruct the 2-D structure of the Be disc by applying the Doppler tomography method to the H$\alpha$ and H$\beta$ emission line profiles, using a rigidly rotating frame with the V/R variability period.
The resulting disc structure reveals non-axisymmetric features, which can be explained by a one-armed perturbation in the Be disc.
It is the first time that an eccentric disc structure is directly detected by using a method other than the interferometric one.
\end{abstract}


\section{Introduction} 
Classical Be stars (Be stars for short) are non-supergiant B-type stars which show or have shown Balmer lines in emission.
The striking feature of the Be stars is their high rotation speeds.
They rotate at a speed close to critical so that the surface gravity mostly balances with their centrifugal force around the equator.
The mass ejected from the photosphere forms an circumstellar envelope, called a Be disc, which is another striking feature of Be stars.
The Be stars exhibit complicated line profiles containing an  absorption component from the photosphere and an emission component from the disc.
Other observed features of the Be stars are IR-excess due to $f$-$f$ and $f$-$b$ emission, and linear polarization due to electron scattering in the Be disc.
The emission line profiles, reflecting the state of the Be disc, show many kinds of variability on the timescales from days to several decades, e.g., disappearance and reformation of the disc, and line profile variability (LPV) including  V/R variability, the variability in the ratio of the intensity of violet (V) and red (R) peaks of double-peaked emission line profiles.

Recently, \citet{PR03} reviewed the latest results of Be stars: mainly their properties, variabilities, and probable mechanisms of the disc formation.
The mechanism of Be-disc formation is still under debate, because it is difficult to model the complex mass ejection process, taking into account the various physics at work, e.g., the effect of rapid rotation, the radiative force, and non-LTE calculations.
The common understanding is, however, that the key to Be-disc formation is the high rotation speed of the Be stars [for further discussion, see \citet{PR03} and references therein].
Some authors have suggested that the high rotation speed is linked to stellar evolution \citep*{McS05,Eks08}.

Be/X-ray binaries are systems which consist of a Be star and a compact object (mostly a neutron star) and account for a large fraction of high-mass X-ray binaries \citep{Coe09}.
The system has two discs: a Be disc and an accretion disc around the neutron star, which is formed by material transferred from the Be disc.
Hence, the Be/X -ray binaries exhibit accretion phenomena as well as Be phenomena.

Generally, the eccentricity of the Be/X-ray binary is not small ($\gtrsim 0.3$), which implies that the interaction between the two stars and the mass transfer from the Be disc to the neutron star depends on the orbital phase \citep{Oka01b}.
Most of the Be/X-ray binaries are thus transient X-ray  sources.
The X-ray outbursts from these systems are divided into two types according to the X-ray luminosity: normal outbursts (type I outbursts; $\sim$ $10^{36 - 37}$ erg s$^{-1}$  at 1 -- 20 keV), and giant outbursts (type II outbursts; " $10^{37}$ erg s$^{-1}$ at 1 -- 20 keV).
The normal outbursts, which last several days, are known to occur around periastron passage of the neutron star.
\citet{Oka01b} applied the resonantly truncated Be-disc model developed by \citet{Neg01} to several Be/X-ray binary systems.
They concluded that the normal outbursts occur in systems with intermediate to high eccentricities where the mass transfer from the Be disc to the neutron star takes place at every periastron passage.
On the other hand, the giant outbursts, lasting several tens of days with no orbital modulation, is not well understood.

A0535+26/V725 Tau, which was discovered by the Ariel V satellite during a giant outburst \citep{Ros75, Coe75}, is one of the best studied Be/X-ray binaries.
The system consists of an X-ray pulsar of 103 seconds spin period \citep{Cab07} and an O9.7IIIe star \citep{Gia80}.
The recurrence time of  the normal outbursts and spin-down rate of the pulsar provide the orbital period of 111 days \citep{Mot91} and the eccentricity of 0.47 \citep{Fin94}.
Neither optical photometry nor spectroscopy, however, has yielded an orbital period consistent with X-ray data \citep*[e.g.,][]{Wan98,Lar01},  because the optical data shows very complex variabilities of line profiles due to the variability of the Be star.

Long-term variabilities in this system have been reported by many authors.
\citet*{Cla98a,Cla98b,Cla99} performed UV, optical and IR spectroscopy and $U$, $B$, $V$-band photometry.
Seven-year optical spectroscopy exhibited quasi-periodic variability with a period of $\sim$ 1 year, and fifteen-year photometry showed variability in phase with the spectroscopic one.
\citet{Hai04} reported cyclic behaviour due to resonant truncation of the Be disc by carrying out optical spectroscopy and $I$, $J$, $K$-band photometry for fourteen years.
The dramatic Be disc growth between 1998 and 2000 was observed by \citet{Gru07}, which also reported V/R variabilities.

However, no short-term variabilities (a few week or less) can be discussed from these observations, since they have not been carried out densely enough.
Besides, many optical spectroscopic observations reported previously had low or medium dispersion, and it is therefore difficult to discuss small amplitude ($\lesssim$ 10 km/s) variability in the disc.  

In this two-paper series, we report on, both of the short-term and long-term variabilities of the Be disc in A0535+26/V725 Tau system, based on optical high-dispersion spectroscopic monitoring observations from November 2005 to March 2009.
In this paper (paper I), we focused on long-term ($\gtrsim$ years) variabilities.
Short-term (from days to weeks) variabilities will be discussed in paper II (Moritani et al. in preparation).

In Sec. 2, we summarise our observations.
The results are given in Sec. 3, and we discuss the long-term variability of A0535+26 in Sec. 4.
We present our conclusions in Sec. 5.

\section{Observation} 
\subsection{configuration}
\begin{table*}
	\caption{Observation log. Observations are divided into nine periods displayed in column 1 (see text for details). Column 3 shows HJD of the mid exposure time of each observation. Orbital phases corresponding to three different ephemerides are listed in column 5 to column 7, while the phase based on the X-ray light curve is given in column 8 (for the definition of each ephemeris, see table \ref{periastron}). 
Columns 9 and 10 show whether the H$\alpha$ and H$\beta$ line profiles respectively were observed ($\bigcirc$) or not (---).}
		\begin{center}
			\begin{tabular}{cccccccccc}\hline \hline
				Observation	&	&	&	& $\phi_1$	& $\phi_2$	& $\phi_3$	& $\phi_X$	&	&	\\
				Period	& Date & HJD	&Observatory  &  (Ephemeris 1)	&  (Ephemeris 2)	&  (Ephemeris 3)	&  (Outburst)	& H$\alpha$	& H$\beta$	\\ \hline
					I	& 2005 Nov. 24	& 2 453 699.060	& OAO	& 0.576 & 0.066 & 0.773 & 0.727	& $\bigcirc$	& ---	\\
						& 2005 Nov. 25	& 2 453 700.060	& OAO	& 0.585 & 0.075 & 0.782 & 0.736	& $\bigcirc$	& --- 	\\
						& 2005 Nov. 27	& 2 453 702.060	& OAO	& 0.603 & 0.093 & 0.801 & 0.754	& $\bigcirc$	& --- 	\\
						& 2005 Nov. 29	& 2 453 704.031	& OAO	& 0.621 & 0.111 & 0.818 & 0.772	& $\bigcirc$	& --- 	\\
						& 2005 Nov. 30	& 2 453 705.081	& OAO	& 0.631 & 0.120 & 0.828 & 0.782	& $\bigcirc$	& --- 	\\
						& 2005 Dec. 01	& 2 453 706.075	& OAO	& 0.639 & 0.129 & 0.837 & 0.791	& $\bigcirc$	& --- 	\\
						& 2005 Dec. 03	& 2 453 708.651	& OAO	& 0.663 & 0.153 & 0.860 & 0.814	& $\bigcirc$	& --- 	\\	\hline
					II	& 2006 Dec. 18	& 2 454 088.100	& OAO	& 0.072 & 0.593 & 0.310 & 0.256	& $\bigcirc$	& --- 	\\	\hline
					III	& 2007 Nov. 07	& 2 454 412.184	& OAO	& 0.984 & 0.531 & 0.256 & 0.196	& $\bigcirc$	& $\bigcirc$	\\
						& 2007 Nov. 08	& 2 454 413.172	& OAO	& 0.993 & 0.540 & 0.265 & 0.205	& $\bigcirc$	& $\bigcirc$	\\
						& 2007 Nov. 09	& 2 454 414.141	& OAO	& 0.001& 0.549 & 0.274 & 0.214	& $\bigcirc$	& $\bigcirc$	\\
						& 2007 Nov. 10	& 2 454 415.232	& OAO	& 0.011 & 0.559 & 0.284 & 0.224	& $\bigcirc$	& ---	\\
						& 2007 Nov. 11	& 2 454 416.166	& OAO	& 0.019 & 0.567 & 0.292 & 0.232	& $\bigcirc$	& $\bigcirc$	\\
						& 2007 Nov. 13	& 2 454 418.159	& OAO	& 0.037 & 0.585 & 0.311 & 0.250	& $\bigcirc$	& $\bigcirc$	\\
						& 2007 Nov. 14	& 2 454 419.238	& OAO	& 0.047 & 0.595 & 0.320 & 0.260	& $\bigcirc$	& $\bigcirc$ 	\\	\hline
					IV	& 2007 Dec. 16	& 2 454 451.168	& GAO	& 0.334 & 0.885 & 0.611 & 0.550	& $\bigcirc$	& $\bigcirc$	\\
						& 2007 Dec. 19	& 2 454 454.294	& GAO	& 0.362 & 0.913 & 0.639 & 0.578	& $\bigcirc$	& $\bigcirc$	\\
						& 2007 Dec. 20	& 2 454 455.264	& GAO	& 0.371 & 0.922 & 0.648 & 0.587	& $\bigcirc$	& $\bigcirc$	\\
						& 2007 Dec. 26	& 2 454 461.090	& GAO	& 0.423 & 0.975 & 0.701 & 0.640	& $\bigcirc$	& $\bigcirc$	\\
						& 2008 Jan. 02	& 2 454 468.022	& OAO	& 0.485 & 0.037 & 0.764 & 0.702	& $\bigcirc$	& $\bigcirc$	\\
						& 2008 Jan. 31	& 2 454 497.004	& OAO	& 0.746 & 0.300 & 0.027 & 0.965	& $\bigcirc$	& $\bigcirc$	\\	\hline
					V	& 2008 Mar. 11	& 2 454 537.002	& GAO	& 0.105 & 0.663 & 0.391 & 0.328	& $\bigcirc$	& ---	\\
						& 2008 Mar. 21	& 2 454 547.023	& OAO	& 0.195 & 0.754 & 0.482 & 0.419	& $\bigcirc$	& $\bigcirc$	\\ \hline
					VI	& 2008 Oct. 01	& 2 454 741.208	& OAO	& 0.940 & 0.514 & 0.247 & 0.180	& $\bigcirc$	& $\bigcirc$	\\
						& 2008 Oct. 11	& 2 454 751.162	& OAO	& 0.029 & 0.604 & 0.338 & 0.271	& $\bigcirc$	& $\bigcirc$	\\
						& 2008 Oct. 11	& 2 454 751.229	& GAO	& 0.030 & 0.605 & 0.338 & 0.271	& $\bigcirc$	& $\bigcirc$	\\
						& 2008 Oct. 15	& 2 454 755.186	& GAO	& 0.065 & 0.641 & 0.374 & 0.307	& $\bigcirc$	& $\bigcirc$	\\
						& 2008 Oct. 19	& 2 454 759.141	& GAO	& 0.101 & 0.677 & 0.410 & 0.343	& $\bigcirc$	& $\bigcirc$	\\
						& 2008 Nov. 04	& 2 454 775.102	& GAO	& 0.244 & 0.821 & 0.555 & 0.488	& $\bigcirc$	& $\bigcirc$	\\
						& 2008 Dec. 12	& 2 454 813.068	& GAO	& 0.586 & 0.166 & 0.901 & 0.832	& $\bigcirc$	& $\bigcirc$	\\ \hline
					VII	& 2008 Dec. 25	& 2 454 826.207	& OAO	& 0.704 & 0.285 & 0.020 & 0.952	& $\bigcirc$	& $\bigcirc$	\\
						& 2008 Dec. 26	& 2 454 827.097	& OAO	& 0.712 & 0.293 & 0.028 & 0.960	& $\bigcirc$	& $\bigcirc$	\\
						& 2008 Dec. 27	& 2 454 828.129	& OAO	& 0.721 & 0.302 & 0.038 & 0.969	& $\bigcirc$	& $\bigcirc$	\\
						& 2008 Dec. 28	& 2 454 829.187	& OAO	& 0.730 & 0.312 & 0.047 & 0.979	& $\bigcirc$	& $\bigcirc$	\\ \hline
					VIII	& 2008 Dec. 29	& 2 454 830.165	& OAO	& 0.739 & 0.321 & 0.056 & 0.987	& $\bigcirc$	& $\bigcirc$	\\
						& 2008 Dec. 30	& 2 454 831.157	& OAO	& 0.748 & 0.330 & 0.065 & 0.996	& $\bigcirc$	& $\bigcirc$	\\
						& 2008 Dec. 31	& 2 454 832.112	& OAO	& 0.757 & 0.338 & 0.074 & 0.005	& $\bigcirc$	& $\bigcirc$	\\
						& 2009 Jan. 01	& 2 454 833.008	& OAO	& 0.765 & 0.346 & 0.082 & 0.013	& $\bigcirc$	& $\bigcirc$	\\
						& 2009 Jan. 02	& 2 454 834.021	& OAO	& 0.774 & 0.356 & 0.091 & 0.022	& $\bigcirc$	& $\bigcirc$	\\
						& 2009 Jan. 03	& 2 454 835.028	& OAO	& 0.783 & 0.365 & 0.100 & 0.032	& $\bigcirc$	& $\bigcirc$	\\
						& 2009 Jan. 05	& 2 454 837.114	& OAO	& 0.802 & 0.384 & 0.119 & 0.050	& $\bigcirc$	& $\bigcirc$	\\
						& 2009 Jan. 06	& 2 454 838.138	& OAO	& 0.811 & 0.393 & 0.129 & 0.060	& $\bigcirc$	& $\bigcirc$	\\
						& 2009 Jan. 07	& 2 454 839.180	& OAO	& 0.820 & 0.402 & 0.138 & 0.069	& $\bigcirc$	& $\bigcirc$	\\
						& 2009 Jan. 08	& 2 454 840.194	& OAO	& 0.829 & 0.412 & 0.147 & 0.078	& $\bigcirc$	& $\bigcirc$	\\
						& 2009 Jan. 10	& 2 454 842.186	& OAO	& 0.847 & 0.430 & 0.165 & 0.096	& $\bigcirc$	& $\bigcirc$	\\
						& 2009 Jan. 12	& 2 454 844.157	& OAO	& 0.865 & 0.447 & 0.183 & 0.114	& $\bigcirc$	& $\bigcirc$	\\ \hline
					IX	& 2009 Mar. 12	& 2 454 902.965	& GAO	& 0.393 &  0.981 &  0.718 &  0.648	& $\bigcirc$	& $\bigcirc$	\\ \hline
				\end{tabular}
	\end{center}
	\label{log}
\end{table*}

We carried out high dispersion optical spectroscopic monitoring observations of A0535+26 from November 2005 to March 2009, mainly at the Okayama Astrophysical Observatory (OAO) with a 188 cm telescope equipped with HIDES (High Dispersion Echelle Spectrograph).
Observations were also performed at Gunma Astronomical Observatory (GAO) with a 1.5 m telescope equipped with GAOES (Gunma Astronomical Observatory Echelle Spectrograph).
HIDES covers a 1200 \AA \ (5500 -- 6700 \AA \ and 4400 -- 5600 \AA ) wavelength range until 2007, and 3500 \AA \ (4000 -- 7500 \AA ) from 2008 thanks to adoption of mosaicked three EEV 42 -- 80 CCDs (2048 $\times$ 4098 $pix^2$).
The wavelength coverage of GAOES is 1900 \AA \ (4800 -- 6700 \AA ) and the detector is a EEV 44 -- 82 CCD.

The typical wavelength resolution $R$ and the signal to noise ratio $S/N$ of our OAO/HIDES data around H$\alpha$ are $R$ $\sim$ 60000 and  $S/N$ $\sim$ 120, respectively.
For H$\beta$, $R$ $\sim$ 60000 and $S/N$ $\sim$ 100.
On the other hand, our H$\alpha$ data obtained with GAO/GAOES has $R$ $\sim$ 30000 and $S/N$ $\sim$ 120, and the H$\beta$ data has  $R$ $\sim$ 30000 and $S/N$ $\lesssim$ 100.
The exposure time for the H$\alpha$ data with OAO/HIDES ranged from 3600 to 5400 s, and that of H$\beta$ data was from 3600 to 7200 s.
With GAO/GAOES the exposure time was 3600 -- 9600 s.
We combined 1200- or 1800- second-exposure data into daily averaged spectra.

The obtained data is reduced in the standard way using IRAF\footnote{IRAF (Imaging Reduction and Analysis Facility) is a software system for the reduction and analysis of astronomical data, supported by NOAO.\\http://iraf.noao.edu/} echelle package -- subtraction of bias, flat fielding, calibration of the wavelength using Th-Ar lines, normalization of the continuum, and correction to the helio-centric orbit.

\subsection{observation periods}
Our observation log is listed in table \ref{log}.
The observation period in column 1 is described below.
From columns 5 to 8 are listed the orbital phases corresponding to three different ephemerides by different authors and the X-ray light curve.
The time of phase 0 ($HJD_{0}$) is given by the following equation,
\begin{equation}
HJD_{0} = HJD_{origin} + P_{orb} \times E,
\end{equation}
where $HJD_{origin}$ is the origin, $P_{orb}$ the orbital period, and E an integer indicating the cycle number.
$HJD_{origin}$ corresponds to a periastron passage in the case of ephemeris, while it corresponds to the time when the X-ray flux starts to rise in the case of X-ray light curve (see below).
We used three different ephemerides reported by different authors because of difficulty in determining the time of periastron passage.
The $HJD_{origin}$ and $P_{orb}$ of each ephemeris are summarised in table \ref{periastron}.

\begin{figure}
	\begin{center}
		\includegraphics[width=80mm]{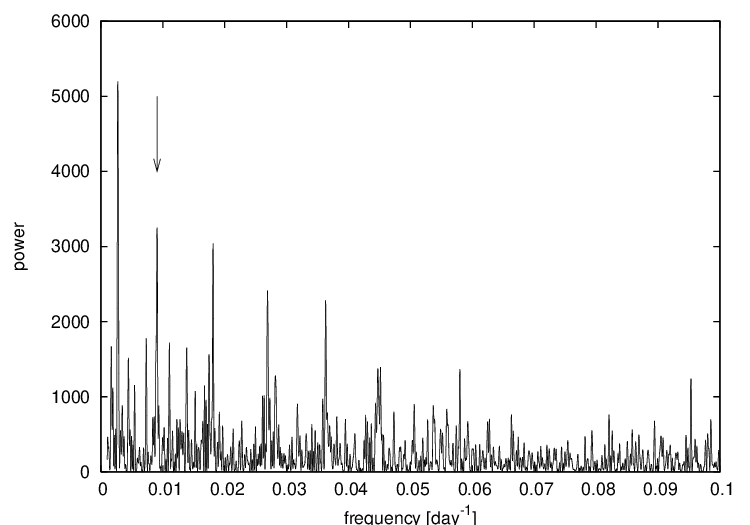}
	\end{center}
	\caption{Power spectrum of the RXTE/ASM data. The arrow in the spectrum indicates the most plausible value, which is corresponding to 110.24 days. 
The highest peak corresponds to 366.45 days, introduced by yearly sun angle constraints in RXTE observations of A0535+26.}
	\label{fourier1}
\end{figure}

\begin{figure}
	\begin{center}
		\includegraphics[width=80mm]{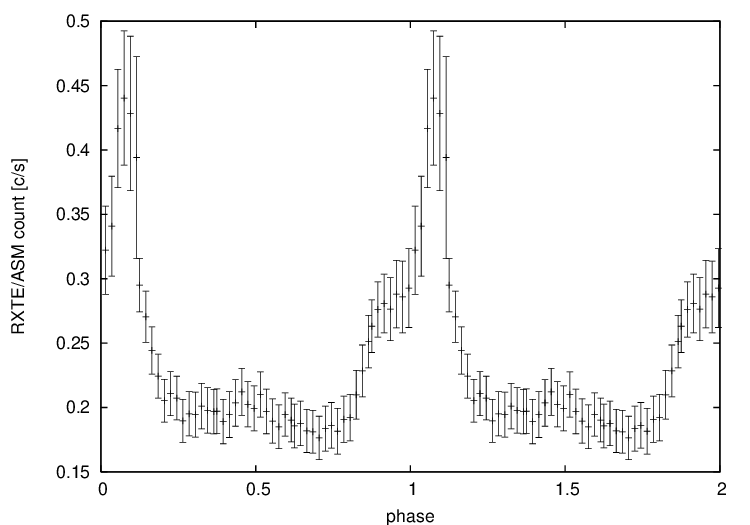}
	\end{center}
	\caption{X-ray light curve folded by assuming $P_{orb}$=110.24 days and $HJD_{origin}$ = 2453398.43, using the n $\times$ m bin method in \citet{Coe06}. The maximal duration of the normal outbursts is $\phi_X$ = 0.05 -- 0.08.}
	\label{rxte2}
\end{figure}

The outburst phase listed in column 8 is obtained by using RXTE/ASM\footnote{http://xte.mit.edu/asmlc/ASM.html} one-day averaged data from JD 2450133 (2 February 1996) to JD 2454881 (19 February 2009) as follows: First, we determine the orbital period by the Fourier analysis of the X-ray light curve.
To remove uncertainty due to giant outbursts, the data between $-$1.5 and 4.0 ASM Unit counts s$^{-1}$ (1 Crab is approximately 75 ASM Unit counts/s at 2 -- 10 keV) were used for our period analysis. 
The power spectrum is shown in figure \ref{fourier1}.
The arrow in the power spectrum indicates the most plausible peak: 0.00907$\pm$0.00005 [day$^{-1}$], corresponding to $P_{orb}$=110.2$\pm$0.6 days.
Other peaks are overtones or harmonics with one year.
Then, the beginning time of an X-ray normal outburst, HJD 2453398.4, is taken as $HJD_{origin}$.
The obtained period and the origin are listed in table \ref{periastron} together with those in the three different ephemerides.
Figure \ref{rxte2} shows the X-ray light curve folded with these $P_{orb}$ and $HJD_{origin}$ by the n $\times$ m bin method used in \citet{Coe06}.
The light curve shows that normal outbursts reach the maximum flux at $\phi_X$ = 0.05 -- 0.08.

\begin{table*}
	\caption{The orbital period, $P_{orb}$, and the origin of the phase, $HJD_{origin}$, in each ephemeris and those derived from the X-ray light curve. See text for detail.}
	\begin{center}
			\begin{tabular}{ccccc}\hline \hline
				Phase	& Type	& $P_{orb}$	& $HJD_{origin}$ 	& references	\\	\hline
					$\phi_1$	& Ephemeris 1	& 111.38$\pm$ 0.11	& 2 446 734.3$\pm$2.6	& \citet{Mot91}		\\
					$\phi_2$	& Ephemeris 2	& 110.3$\pm$0.3	& 2 449 059.2$\pm$0.6	& \citet{Fin94}		\\
					$\phi_3$	& Ephemeris 3	& 110.0$\pm$0.5	& 2 450 094$\pm$1	& \citet{Coe06}	\\
					$\phi_X$	& Outburst	& 110.2$\pm$0.6	& 	2 453 398.4$\pm$1.0 	& this work	\\ \hline
				\end{tabular}
	\end{center}
	\label{periastron}
\end{table*}

Our monitoring term is divided into nine periods (as shown in column 1 of table \ref{log}).
We carried out the monitoring observations, mainly for detecting the variability of the Be disc around periastron passage, when the Balmer line is expected to change due to the tidal interaction between the Be disc and the neutron star, but also for examining long-term variabilities such as V/R variations.
Observations in periods I (in November 2005), III (in November 2007) and VIII (in January 2009), around periastron passage with respect to the $\phi_2$, $\phi_1$ and $\phi_X$ ($\phi_3$), respectively, are performed densely, aiming for studying short-term variability.
Observations in other periods are carried out in order to monitor A0535+26 between normal outbursts as well as to aim at long-term variability. 

Recently, \citet{Lev08} reported that A0535+26 brightened in X-rays on 12 September 2008 (JD 2454721) at 21 mCrab at 2 -- 10 keV, and that the flux then rose up to approximately 48 mCrab (3.6 ASM Unit counts s$^{-1}$).
Observations in the period VI, from 1 October to 4 November in 2008, was carried out thanks to the ToO (Target of Opportunity) observation program at OAO, in order to examine the variability of the Be disc at $\phi_X$ = 0.2 -- 0.5 after the normal outburst.

Based on the ephemeris given by \citet{Coe06}, the observation of periods VII and VIII were scheduled; the next periastron passage was predicted to be around 31 December 2008, since the last outburst occurred on 12 September after periastron.
The activity reports of the RXTE/ASM team declared that normal outburst occurred in the week of 2 -- 9 January 2009; the X-ray photon count started to rise on JD 2454832.
Our observations were performed from 25 December 2008 thorough 12 January 2009, which is divided into two periods; pre-outburst (period VII, until 28 December) and around outburst (period VIII, from 29 December). 
An additional observation was made at GAO on 12 March 2009 (in the period IX).

\section{Results} 
\begin{figure}
	\begin{center}
		\includegraphics[width=72mm]{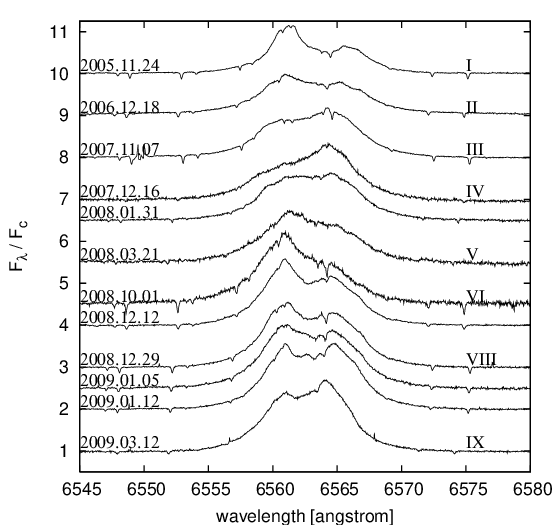}
	\end{center}
	\caption{Representative H$\alpha$ line profiles from our monitoring observations. The observation date is annotated on each spectrum. 
There are offsets between each spectrum for clarity, and larger space is inserted between spectra in different periods. Spectra in period VII are not shown since they were similar to that on 29 December 2008. 
The line profile variability is clearly seen. See text for detail.
}
	\label{long_halpha}
\end{figure}

\begin{figure}
	\begin{center}
		\includegraphics[width=72mm]{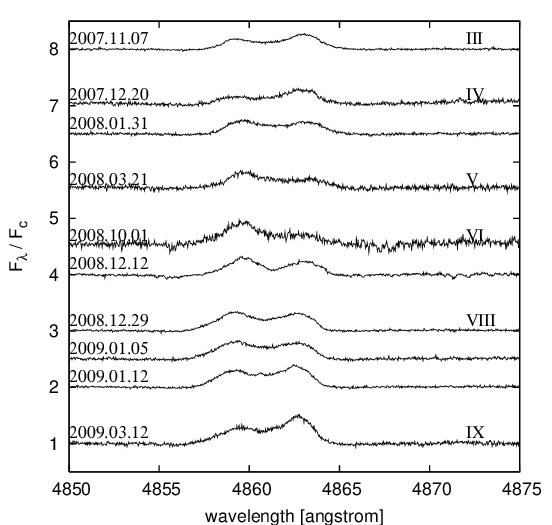}
	\end{center}
	\caption{Representative H$\beta$ line profiles obtained from our monitoring observations. As in Fig. \ref{long_halpha}, the observed date is annotated on each spectrum, and a larger space is inserted between spectra in different periods. 
The H$\beta$ line has kept double-peaked, but the V/R ratio had variabilities nearly in phase with H$\alpha$ line profile (see Fig. \ref{VR_both} and text).}
	\label{long_hbeta}
\end{figure}

Figure \ref{long_halpha} and \ref{long_hbeta} show representative spectra of H$\alpha$ and H$\beta$ obtained during the monitoring observations; all spectra will be shown in paper II.
For reasons of clarity, the spectra are shown on the same Y-scale and with linear offsets from each other.
For three and a half years, both H$\alpha$ and H$\beta$ line profiles have been in emission, which indicates that the Be disc of A0535+26 did not disappear.
The profiles, however, have changed.
In period I, the H$\alpha$ line profile was double-peaked with the violet component stronger than the red component (V $>$ R, the top profile in figure \ref{long_halpha}).
One year later (second spectrum from the top), in December 2006, the profile kept V $>$ R, but the V/R ratio has decreased from $>$ 1.7 to 1.22.
Here, the V/R ratio was determined by smoothing the spectra and measuring the peak of the intensities normalized by the continuum.
It turned to V $<$ R (third spectrum from the top) in period III (in November 2007).
In period IV, the H$\alpha$ line changed from a red-enhanced single-peaked profile to a single-peaked profile with a nearly flat top in less than six weeks (fourth and fifth spectra from the top), followed by the double-peaked profile with V $>$ R after two months (in March 2008, beneath the two).
From October 2008 to March 2009, from period VI to IX, the H$\alpha$ line profile remained double-peaked, gradually changing from V $>$ R to V $<$ R in more than half a year (sixth spectrum from the bottom).
Besides, in period VIII, the profile showed an obvious variability within less than two weeks.
This short-term variability, which we think is due to the tidal interaction between the Be disc and the neutron star, will be discussed in detail in paper II.

\begin{figure}
	\begin{center}
		\includegraphics[width=80mm]{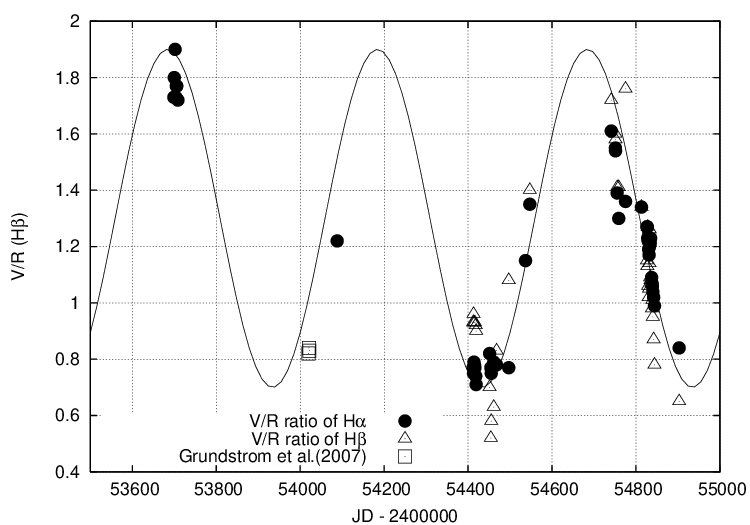}
	\end{center}
	\caption{V/R ratio of the H$\alpha$ ({\it filled circles}) and H$\beta$ ({\it triangles}) line profiles vs. HJD in our observation in periods I -- IX. A remarkable similarity is seen in the variation patterns of both lines.
The V/R ratio in \citet{Gru07} is also plotted (squares).
The solid line is a sine curve with the period of 500 days (see also Fig. \ref{VR_fold}).
}
	\label{VR_both}
\end{figure}

As shown in figure \ref{long_hbeta}, the H$\beta$ line profile was always double-peaked as far as our observations were carried out, but the V/R ratio exhibited variabilities.
In period III, the red component was stronger than the violet, i.e. V $<$ R.
In periods IV and V (from December 2007 to March 2008), the V/R ratio of H$\beta$ line changed from V $<$ R to V $>$ R within a couple of months.
Then, the V/R ratio gradually turned again to V $>$ R during the following year.
The variability of the V/R ratio is almost in phase with that of the H$\alpha$ line profile (see table \ref{result_obs} and figure \ref{VR_both}).

The obtained EW and V/R ratio of all the H$\alpha$ and H$\beta$ line profiles are listed in table \ref{result_obs}.
The typical error in the value of EW is $\sim$ $\pm$ 0.05 \AA \ , and that of V/R ratio is $\pm$ 0.03.
In period IV, the V/R ratio of H$\alpha$ line profiles could not be obtained because the line profiles did not have double peak.
In the long term, both EW (H$\alpha$) and EW (H$\beta$) gradually increased in amplitude during the last three and a half years, which indicates that the Be disc of A0535+26 has become more and more active.

\begin{table*}
	\caption{List of the EW and V/R ratio of all the H$\alpha$ and H$\beta$ emission line profiles observed from November 2005 to March 2009. Column 1 and 2 list the observation date and corresponding HJD. The orbital phase based on the X-ray light curve, $\phi_X$ (column 8 in table \ref{log}), is listed again in column 3. In columns 4 and 5, EW and V/R ratio of H$\alpha$ are listed, while in column 6 and 7 are listed those of H$\beta$. The typical errors in EW and V/R ratio are $\sim$ $\pm$ 0.05 \AA , and $\pm$ 0.03, respectively.}
		\begin{minipage}{120mm}
			\begin{tabular}{ccccccc}\hline \hline
				Day	& HJD	& $\phi_X$	& EW (H$\alpha$) 	& V/R (H$\alpha$)	& EW (H$\beta$)	& V/R (H$\beta$)	 \\ \hline
					2005 Nov. 24	& 2 453 699.060	& 0.727	& $-$6.64\AA \  & 1.73 & --	& --	\\
					2005 Nov. 25	& 2 453 700.060	& 0.736	& $-$6.72 \AA \ & 1.81 & --	& --	\\
					2005 Nov. 27	& 2 453 702.060	& 0.754	& $-$6.73 \AA \ & 1.90 & --	& --	\\
					2005 Nov. 29	& 2 453 704.031	& 0.772	& $-$6.68 \AA \ & 1.74 & --	& --	\\
					2005 Nov. 30	& 2 453 705.081	& 0.782	& $-$6.85 \AA \ & 1.80 & --	& --	\\
					2005 Dec. 01	& 2 453 706.075	& 0.791	& $-$6.49 \AA \ & 1.80 & --	& --	\\
					2005 Dec. 03	& 2 453 708.051	& 0.814	& $-$6.45 \AA \ & 1.73 & --	& --	\\	\hline
					2006 Dec. 18	& 2 454 088.100	& 0.256	& $-$9.84 \AA \ & 1.22 & --	& --	\\	\hline
					2007 Nov. 07	& 2 454 412.184	& 0.196	& $-$9.74 \AA \ & 0.77 & $-$0.95 \AA \ & 0.94	\\
					2007 Nov. 08	& 2 454 413.172	& 0.205	& $-$8.83 \AA \ & 0.76 & $-$0.84 \AA \ & 0.96	\\
					2007 Nov. 09	& 2 454 414.141	& 0.214	& $-$9.42 \AA \ & 0.79 & $-$0.85 \AA \ & 0.93 	\\
					2007 Nov. 10	& 2 454 415.232	& 0.224	& $-$9.05 \AA \ & 0.78 & --	& --	\\
					2007 Nov. 11	& 2 454 416.166	& 0.232	& $-$9.55 \AA \ & 0.78 & $-$0.92 \AA \ & 0.93	\\
					2007 Nov. 13	& 2 454 418.159	& 0.250	& $-$9.31 \AA \ & 0.75 & $-$0.97 \AA \ & 0.93	\\
					2007 Nov. 14	& 2 454 419.238	& 0.260	& $-$9.25 \AA \ & 0.72 & $-$0.99 \AA \ & 0.91	\\	\hline
					2007 Dec. 16	& 2 454 451.168	& 0.550	& $-$10.45 \AA \ & *	& $-$1.23 \AA \ 	& 0.70	\\
					2007 Dec. 19	& 2 454 454.294	& 0.578	& $-$9.73 \AA \ & *	& $-$1.28 \AA \ 	& 0.53  	\\
					2007 Dec. 20	& 2 454 455.264	& 0.587	& $-$9.01 \AA \ & *	& $-$0.95 \AA \ 	& 0.58	\\
					2007 Dec. 26	& 2 454 461.090	& 0.640	& $-$9.32 \AA \ & *	& $-$1.06 \AA \ 	& 0.64 	\\
					2008 Jan. 02	& 2 454 468.022	& 0.702	& $-$10.23 \AA \ & *	& $-$1.14 \AA \ 	& 0.84	\\
					2008 Jan. 31	& 2 454 497.004	& 0.965	& $-$10.27 \AA \ & **	& $-$1.32 \AA \ 	& 1.01	\\	\hline
					2008 Mar. 11	& 2 454 537.002	& 0.328	& $-$10.66 \AA \ & 1.15	& --	& --	\\
					2008 Mar. 21	& 2 454 547.023	& 0.419	& $-$9.90 \AA \ &	 1.36	& $-$0.93	&	1.40	\\ \hline
					2008 Oct. 01	& 2 454 741.208	& 0.180	& $-$11.34 \AA \ & 1.61	& $-$1.22 \AA \ & 1.73	\\
					2008 Oct. 11	& 2 454 751.162	& 0.271	& $-$10.20 \AA \ & 1.56	& $-$1.40 \AA \ & 1.61	\\
					2008 Oct. 11	& 2 454 751.229	& 0.271	& $-$10.51 \AA \ & 1.54	& $-$1.27 \AA \ & 1.58	\\
					2008 Oct. 15	& 2 454 755.186	& 0.307	& $-$10.92 \AA \ & 1.39	& $-$1.08 \AA \ & 1.42	\\
					2008 Oct. 19	& 2 454 759.141	& 0.343	& $-$11.24 \AA \ & 1.30	& $-$1.08 \AA \ & 1.41	\\
					2008 Nov. 04	& 2 454 775.102	& 0.488	& $-$11.83 \AA \ & 1.37	& $-$1.26 \AA \ & 1.76	\\
					2008 Dec. 12	& 2 454 813.068	& 0.832	& $-$11.01 \AA \ & 1.35	& $-$1.03 \AA \ & 1.34	\\ \hline
					2008 Dec. 25	& 2 454 826.207	& 0.952	& $-$11.82 \AA \ & 1.27	& $-$1.38 \AA \ & 1.16	\\
					2008 Dec. 26	& 2 454 827.097	& 0.960	& $-$12.02 \AA \ & 1.28	& $-$1.32 \AA \ & 1.14	\\
					2008 Dec. 27	& 2 454 828.129	& 0.969	& $-$12.39 \AA \ & 1.23	& $-$1.36 \AA \ & 1.28	\\
					2008 Dec. 28	& 2 454 829.187	& 0.979	& $-$12.14 \AA \ & 1.22	& $-$1.45 \AA \ & 1.07	\\ \hline
					2008 Dec. 29	& 2 454 830.165	& 0.987	& $-$12.48 \AA \ & 1.20	& $-$1.39 \AA \ & 1.03	\\
					2008 Dec. 30	& 2 454 831.157	& 0.996	& $-$12.44 \AA \ & 1.18	& $-$1.24 \AA \ & 1.06	\\
					2008 Dec. 31	& 2 454 832.112	& 0.005	& $-$12.21 \AA \ & 1.21	& $-$1.42 \AA \ & 1.15	\\
					2009 Jan. 01	& 2 454 833.008	& 0.013	& $-$12.12 \AA \ & 1.22	& $-$1.35 \AA \ & 1.26	\\
					2009 Jan. 02	& 2 454 834.021	& 0.022	& $-$11.74 \AA \ & 1.22	& $-$1.28 \AA \ & 1.09	\\
					2009 Jan. 03	& 2 454 835.028	& 0.032	& $-$11.76 \AA \ & 1.23	& $-$1.47 \AA \ & 1.08	\\
					2009 Jan. 05	& 2 454 837.114	& 0.050	& $-$12.86 \AA \ & 1.10	& $-$1.41 \AA \ & 1.07	\\
					2009 Jan. 06	& 2 454 838.138	& 0.060	& $-$11.67 \AA \ & 1.07	& $-$1.32 \AA \ & 0.98	\\
					2009 Jan. 07	& 2 454 839.180	& 0.069	& $-$12.46 \AA \ & 1.05	& $-$1.54 \AA \ & 1.01	\\
					2009 Jan. 08	& 2 454 840.194	& 0.078	& $-$11.45 \AA \ & 1.05	& $-$1.25 \AA \ & 0.96	\\
					2009 Jan. 10	& 2 454 842.186	& 0.096	& $-$12.19 \AA \ & 1.02	& $-$1.32 \AA \ & 0.88	\\
					2009 Jan. 12	& 2 454 844.157	& 0.114	& $-$12.91 \AA \ & 1.00	& $-$1.47 \AA \ & 0.78	\\ \hline
					2009 Mar. 12	& 2 454 902.965	& 0.648	& $-$12.61 \AA \ & 0.84	& $-$1.74 \AA \ & 0.65	\\ \hline
			\end{tabular}
			* : Red-enhanced profile

			** : Single-peaked profile
		\end{minipage}
	\label{result_obs}
\end{table*}

\section{Discussion} 
The observed H$\alpha$ and H$\beta$ emission-line profiles showed variabilities on timescales from weeks to years. 
In what follows, we focus only on long-term variabilities ($\gtrsim$ year).
Below, we adopt the orbital ephemeris $\phi_X$ based on the X-ray light curve.

\subsection{The  V/R Variation}
\begin{figure}
	\begin{center}
		\includegraphics[width=80mm]{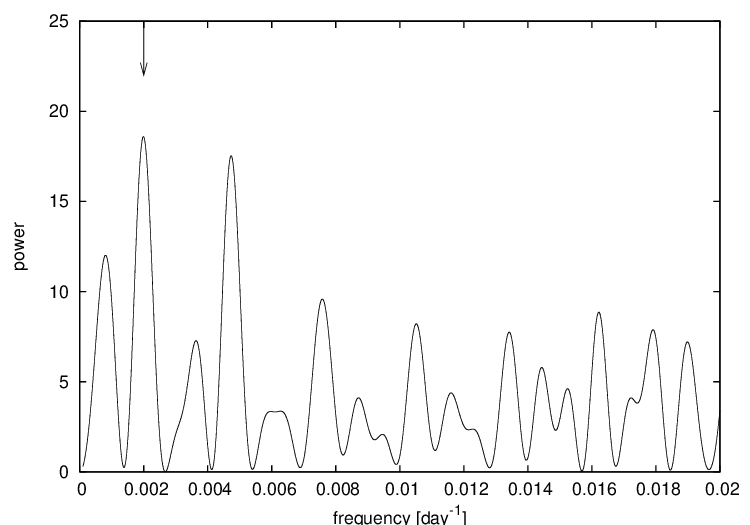}
	\end{center}
	\caption{Power spectrum of the V/R ratio using the data from HJD 2453500 to 2455000. The arrow in the spectrum indicates the most plausible value, which corresponds to 500 days.}
	\label{fourier2}
\end{figure}

\begin{figure}
	\begin{center}
		\includegraphics[width=80mm]{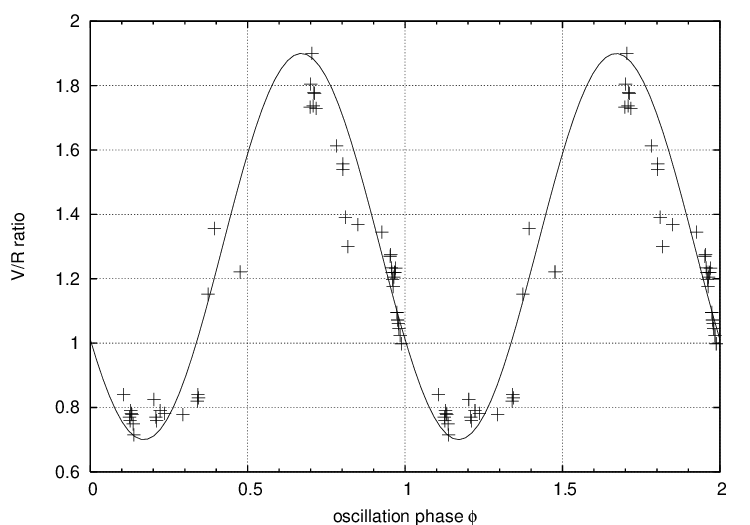}
	\end{center}
	\caption{V/R. variation of H$\alpha$ since 2005 \citep[][and our observations]{Gru07}. The data are folded on 500-day periodicity. The phase 0 is defined as the time when the V/R ratio turns from $>$ 1 to $<$ 1. 
The solid line is a simple sine curve with the same period.
}
	\label{VR_fold}
\end{figure}

As mentioned in the previous section,the relative strength of the violet component of H$\alpha$ compared to the red component was greater than 1 in November 2005 and in December 2006.
Figure 2 of \citet{Gru07}, however, shows that V/R $<$ 1 in October 2006 (square marks in Fig. \ref{VR_both}).
These results suggest that the V/R ratio turned from $>$ 1 to $<$ 1 and then back to $>$ 1 again in 2006 (around JD 2454020).
The V/R ratio also turned from $<$ 1 to $>$ 1 in February 2008 between periods IV and V (around JD 2454500).
It then became $<$ 1 again  between periods VIII and IX (around JD 2454850).
This implies a (quasi-)periodic variation of the V/R ratio, which may be induced by a one-armed oscillation \citep*{Oka91,Pap92}.

We examine the long-term variation of the V/R ratio and EW of H$\alpha$ line from 1979 to 2009 and that of H$\beta$ from 1975 to 2009, using the data from this work, with those available from previous studies [\citet{Hut78}, \citet{Aab85}, \citet{Cla98a} , \citet{Hai04}, \citet{Coe06}, and \citet{Gru07}].
The V/R ratio of the H$\alpha$ seems to have been varying in the range between 0.7 to 2, while that of H$\beta$ between 0.5 and 1.9.
The V/R ratios of both lines are almost in phase, although these lines were seldom observed simultaneously in the previous studies.
We apply the Fourier analysis for the determination of the period of the V/R variation. 
Here, we restrict our analysis to the data of H$\alpha$ line profiles obtained between HJD 2453500 to HJD 2455000, by \citet{Gru07} and this work, because the data are dense enough to derive a reliable period.
Observations before the 2005 giant outburst, which include dense observations from HJD 2451000 to HJD 2452000 \citep*{Gru07,Coe06,Cla98a}, are excluded from the analysis, since it is quite possible that the Be disc have reformed after the outburst.

The period of the V/R variation is determined to be 500$\pm$15 days (figure \ref{fourier2}) at the 90 \% confidence level via $\chi^2$ test, which is significantly longer than that reported in \citet{Cla98a} ($\sim$ 1 year). 
The V/R ratios of the H$\alpha$ spectra from HJD $\sim$2453500 to HJD $\sim$2455000 folded on 500 days of period are plotted in figure \ref{VR_fold}, where the origin of the V/R variation  phase $\Phi_{V/R}$ is taken at the epoch when the V/R ratio turns from $>$ 1 to $<$ 1.
We note that a cyclic behaviour of the V/R ratio is clearly seen  with this period.
The duration when V $>$ R is longer than that for V $<$ R.
The simple sine curve with this period is also drawn in figure \ref{VR_fold} as well as in figure \ref{VR_both}.
\citet{Rei05} studied the period of V/R variations for several Be/X-ray binaries (in their table 3).
For A0535+26/V725 Tau, the V/R period was suggested to be 1 -- 1.5 years \citep*{Hai04,Cla98a}, which is consistent with ours.
We need further observations for a more accurate period, especially between $\Phi_{V/R}$ = 0.4 and $\Phi_{V/R}$ = 0.7 (see figure \ref{VR_fold}).

\citet{Okt09} studied one-armed oscillations of Be disc in circular binaries, taking into account a three-dimensional effect, which \citet{Ogi08} found provides an important contribution to the confinement of the oscillations to the inner part of the Be disc.
They calculated that the period of the fundamental mode of the one-armed oscillation in a B0V-type Be star ranges from 1.4 years to 2.3 years, which depends on the binary separation $D$.
Given that the mass $M_1$ and radius $R_1$ of the Be star V725 Tau are respectively 14 $M_{\odot}$ and 15 $R_{\odot}$ \citep{Gia80}, and that the mass $M_2$ is typical as a neutron star ($\sim$ 1 $M_{\odot}$), $D/R_1$ is approximately 17 in A0535+26, where D is estimated by the Kepler's third law,
\begin{equation}
D^3 = \frac{G(M_1+M_2)}{4\pi ^2}P_{orb}^2.
\end{equation}
If we neglect that the luminosity class of the Be star in A0535+26 is different than in their calculations, we see from figure 3 of \citet{Okt09} that the period of the one-armed oscillation in this system is $\sim$ 2 years, which roughly agrees with our result.
However, the period of a one-armed oscillation is, in general, sensitive to small changes in stellar and disc parameters, and estimated uncertainty of the oscillation period is approximately a factor of two.
More sophisticated calculations, therefore, are needed for further comparison with the observation.

\begin{figure*}
	\begin{center}
		\includegraphics[width=120mm]{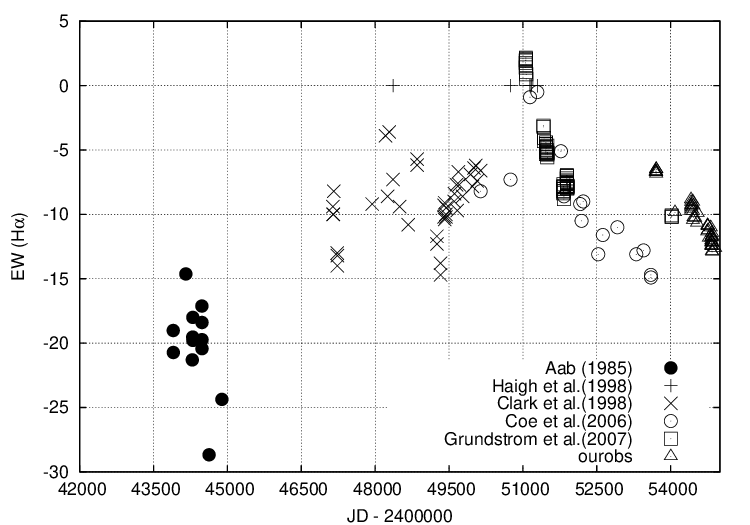}
	\end{center}
	\begin{center}
		\includegraphics[width=120mm]{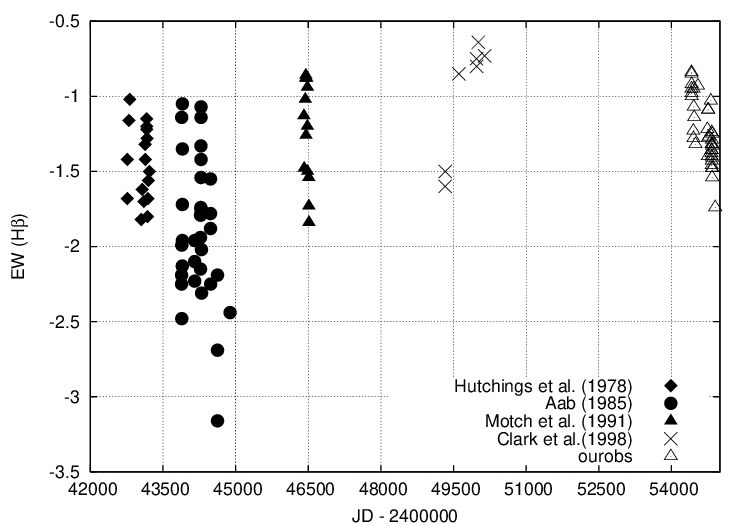}
	\end{center}
	\caption{Long-term variations of EW of the H$\alpha$ ({\it upper panel}) and H$\beta$ ({\it lower panel}) lines
: \citet{Hut78} ({\it diamonds}), \citet{Aab85} ({\it filled circles}), \citet{Mot91} ({\it filled triangles}), \citet{Cla98a} ({\it crosses}), \citet{Hai04} ({\it pluses}) , \citet{Coe06} ({\it open circles}) \citet{Gru07} ({\it squares}), and this work ({\it open triangles}).
}
	\label{EW_history}
\end{figure*}

The variation of EW(H$\alpha$) and EW(H$\beta$) is given in figure \ref{EW_history}.
The H$\alpha$ line has been in emission for most of the  last thirty years; only in 1994, it was in absorption \citep{Gru07}.
The growth of the Be disc is correlated with EW.
The larger values of $|$EW$|$ correspond to denser and/or larger Be discs.
Regarding the last ten years, a giant outburst occurred in 2005 \citep[JD=2453507,][]{Tue05}, around which the emission from the Be disc decreased.
Afterwards, the Be disc has kept growing.

\subsection{The  Structure of the Be Disc}
\begin{figure*}
	\begin{minipage}{0.4\hsize}
	\begin{center}
		\includegraphics[width=72mm]{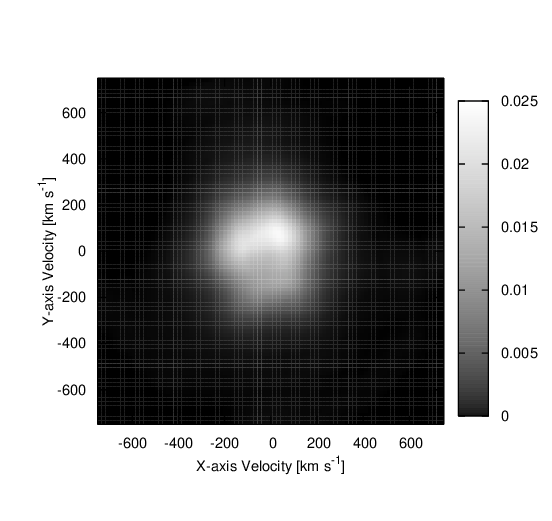}
	\end{center}
	\end{minipage}
	\begin{minipage}{0.4\hsize}
	\begin{center}
		\includegraphics[width=72mm]{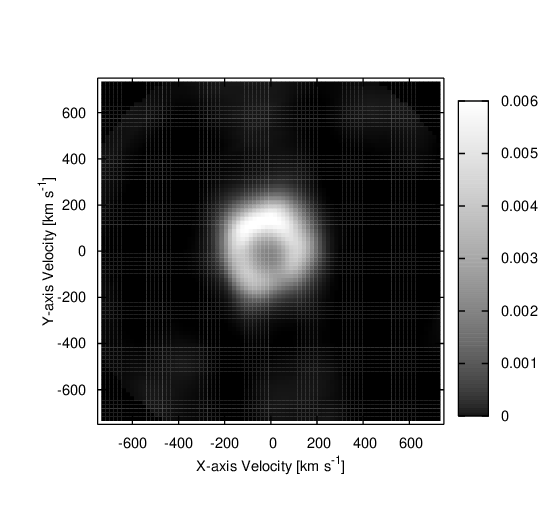}
	\end{center}
	\end{minipage}
	\begin{minipage}{0.4\hsize}
\vspace{-3em}
	\begin{center}
		\includegraphics[width=80mm]{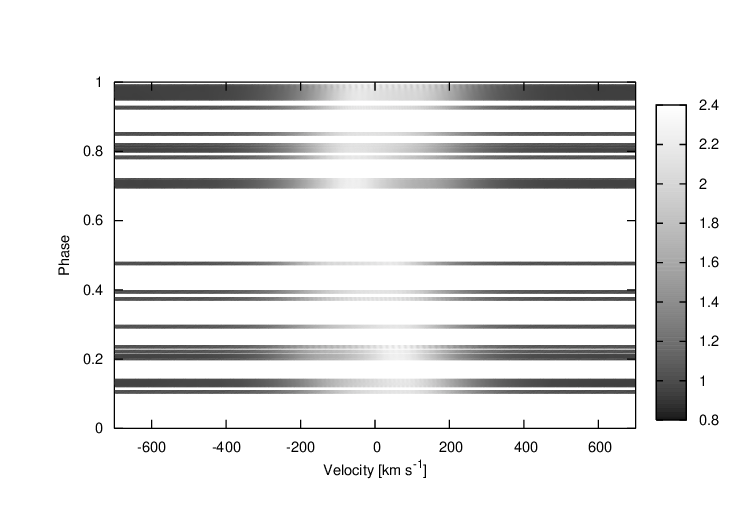}
	\end{center}
	\end{minipage}
	\begin{minipage}{0.4\hsize}
\vspace{-3em}
	\begin{center}
		\includegraphics[width=80mm]{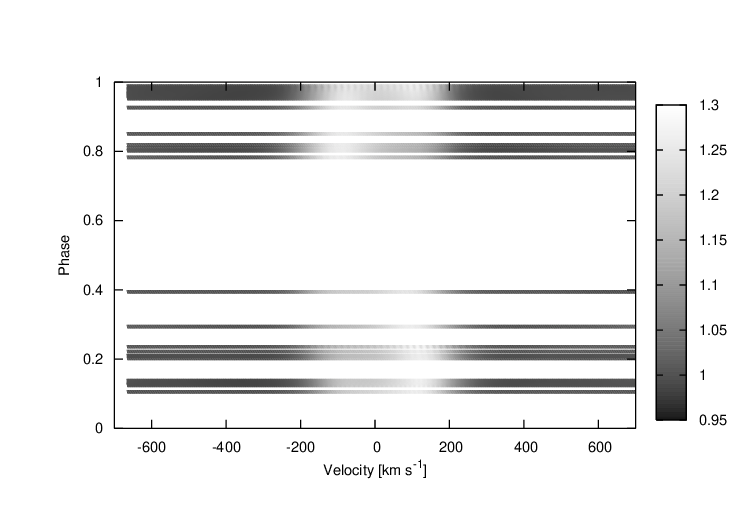}
	\end{center}
	\end{minipage}
	\begin{minipage}{0.4\hsize}
\vspace{-3em}
	\begin{center}
		\includegraphics[width=80mm]{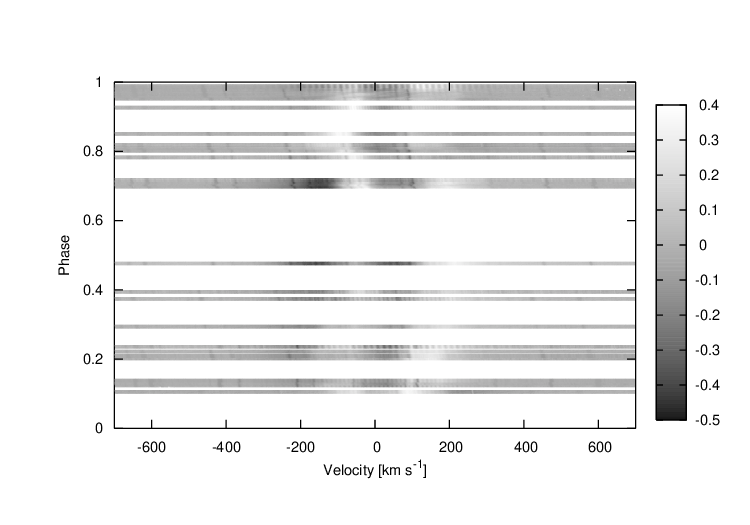}
	\end{center}
	\end{minipage}
	\begin{minipage}{0.4\hsize}
\vspace{-3em}
	\begin{center}
		\includegraphics[width=80mm]{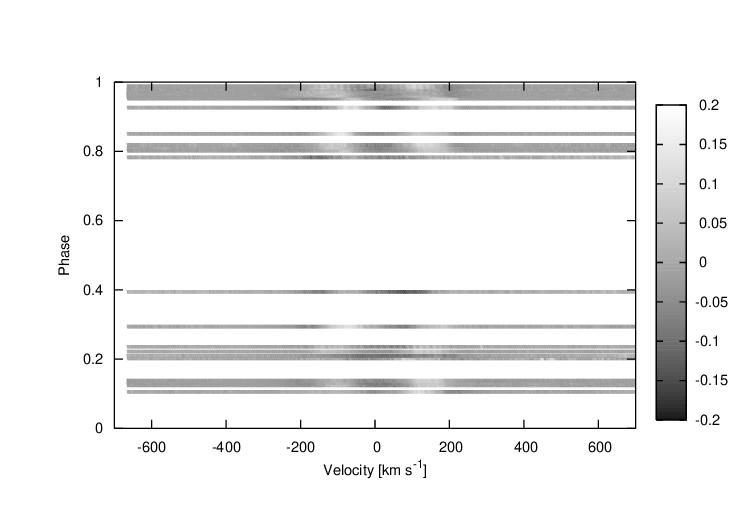}
	\end{center}
	\end{minipage}
	\caption{Results from the Doppler tomography method applied to the H$\alpha$ ({\it left}) and H$\beta$ ({\it right}) line profiles. {\it Top panel}: Obtained disc structure in the velocity plane. The Unit of the structure is arbitrary. {\it Middle panel}: Reconstructed spectrum derived from obtained disc structure. {\it Bottom panel}: Deviation of the reconstructed spectrum from the observed spectrum. The Doppler tomography maps suggest that a non-axisymmetric structure exists in the Be disc of A0535+26.}
	\label{dmap}
\end{figure*}

In order to investigate the two-dimensional structure of the Be disc in A0535+26, we carried out the Doppler tomography method using H$\alpha$ and H$\beta$ spectra.
The Doppler tomography method translates LPV into a distribution map of the emission component.
That is, LPV along the periodic phase (usually the orbital phase) can be projected on a 2-D velocity plane ($V_x$ , $V_y$) corresponding to the Doppler shift of the line profile.
This method is often used for circular binaries in order to obtain the 2-D structure of an accretion disc \citep[see][]{Mar05}.

The Doppler tomography method has been used by several authors for the orbital solution in binary systems.
\citet{Bag94} applied the Doppler tomography to a Be star (29 CMa in a binary system) for the first time.
They used the tomography in order to produce separate UV spectra of the two stars and to obtain their spectral classifications, mass ratio and so forth.
Recently, \citet{Pet08} applied the Doppler tomography method for reconstruction of UV spectra of the secondary FW CMa, because many authors had suggested that its variability in optical and UV spectra was due to the binary interaction although its binarity had been unknown.
They also found that H$\alpha$ and He I $\lambda$6678 variations with its orbital period imply a hot region in the outer area of the Be disc of the primary, facing the secondary.
However, there has never been a direct application of the tomography to the variability of the Be disc.

We have used public software for the Doppler tomography method developed by Henk Spruit\footnote{http://www.mpa-garching.mpg.de/$\tilde{\ }$henk/}.
The V/R variation phase $\Phi_{V/R}$ (determined by the 500-day period) is adopted  instead of the usually-assumed orbital phase, because it is not the binary motion but a one-armed density wave that rotates rigidly.
Our data for the Doppler tomography consists of 47 H$\alpha$ spectra (HJD 2453699.060 -- HJD 2454902.965) and 37 H$\beta$ spectra (HJD 2454412.184	-- HJD 2454902.965).
The coverage of the V/R variation is more than 2 cycles for the H$\alpha$ lines, while $\lesssim$ 1 cycle for the H$\beta$ lines.
The systemic velocity of $-$30 kms$^{-1}$ \citep{Hut84} is taken into account.
However, the radial velocity of the Be star was not subtracted from the result, because it is very difficult to obtain it from each spectrum.
Therefore, each line profile contains an uncertainty in wavelength.
The absolute value of reported radial velocity ranges from several to several tens of  kms$^{-1}$ \citep{Hut78, Gru07}, so that each line profile contains an uncertainty of $\sim$ 0.1 -- 1 \AA \ in wavelength. 
This does not affect the resulting map much (see below).

Figure \ref{dmap} shows results of the Doppler tomography.
The results for the H$\alpha$ line are displayed in the left column, and those for the H$\beta$ line in the right column.
Top panel displays the obtained disc structures with the intensity in arbitrary unit in the velocity plane. 
At $\Phi_{V/R}$ = 0, the observer is in the $+x$-direction, and rotates clockwise.
Obtained spectrum at that time is a projection of Doppler map along the $V_x$ axis.
Note that the plus/minus sign of of $V_x$ and $V_y$ is defined in the x-y plane, and that if a structure, therefore, has a velocity in the $-V_x$-direction, the observer obtains the spectrum with the redshift.
The reconstructed spectra derived from the obtained disc structure and its deviations from the observed ones are shown in the middle and the bottom panels, respectively.
The derived Doppler map shows a non-axisymmetric structure of the Be disc; a brighter region is in $V_y$ $>$ 0, while a fainter region in $V_y$ $<$ 0.
This feature is seen more clearly in the map of H$\beta$ than in H$\alpha$.
These maps suggest the presence of a global one-armed density wave.
Thus, our result provides a direct evidence of the one-armed oscillation in the Be disc.
Interferometric observations have already obtained such density enhancements in the bright Be stars [\citet{Vak98} for $\zeta$ Tau and \citet{Ber99} for $\gamma$ Cas, both with GI2T; see also recent VLTI observations of $\zeta$ Tau \citep{Ste09} and its theoretical modeling \citep{Car09}].
However, this is the first time that a method other than the interferometry directly probed the two-dimensional distribution of the Be disc emissivity. 

There is 10 -- 20 \% difference between observed spectra and reconstructed ones.
We should note here that 1) the radial velocity variations are not corrected in this analysis, as mentioned above, 2) the data are not sufficient especially in $\Phi_{V/R}$ = 0.4 -- 0.7, and 3) the non-axisymmetric structure due to the one-armed oscillation is possibly deformed even within one cycle.
These may affect the resulting Doppler map to some extent.
The first two points should be fixed by future observations, and more theoretical work should be done on the last point.

In the case of isolated Be stars or binary Be stars with well-determined orbital elements, the structure revolving with the period of V/R variation is seen more clearly than the current analysis.
Hence, we can investigate whether the Doppler map contains a perturbation component.
Subtracting the axisymmetric component from the Doppler map, we can obtain the perturbation component of the Be disc.
Then, we can extend the investigation into the question that 1) the perturbation pattern is eccentric or spiral, and 2) one-armed oscillation is confined to inner part of the Be disc.
If the oscillation is confined to the inner part of the disc, the Doppler map in the velocity plane shows an axisymmetric structure only in the inner part, and the outer area of the map is non-axisymmetric.
These studies will be achieved with the future work.

As described above,  the Doppler tomography method is very useful for examining the structure of the Be disc, although the information on time variability cannot be obtained unlike in interferometric observations.
The current study showed that, for Be stars with the periodic V/R variability the 2-D disc structure can be obtained by applying the Doppler tomography method to high-dispersion spectra with 2-m class telescopes.

\section{Conclusions}
We have carried out high-dispersion spectroscopic monitoring observations of the Be/X-ray binary A0535+26/V725 Tau from November 2005 to March 2009.
Regarding the long-term variability, our results are summarised as follows:
\begin{enumerate}
	\item The orbital period is determined to be 110.2 days by the Fourier analysis of RXTE/ASM data.
	\item The EW(H$\alpha$) and EW(H$\beta$) indicate that the Be disc in A0535+26 has kept growing after the last giant outburst in 2005.
	\item From our observations, together with those by \citet{Gru07}, the period of the V/R variations is determined to be approximately 500 days, which is consistent with \citet{Rei05}.
	\item The Doppler tomography method applied to the H$\alpha$ and H$\beta$ line profiles has revealed a non-axisymmetric disc structure precessing with the V/R period, which can be originated from the global one-armed oscillation.
\end{enumerate}
Further observation is needed to improve these results.
In particular, the observations of metallic absorption lines between 4000 and 5000 \AA \ is very important in order to obtain the ephemeris and then the radial velocity of the Be star.

\vspace{3em}
We are very grateful to Ryuko Hirata for his extensive of advice, and to Bun'ei Sato for kindly observing A0535+26.
This paper is based on observations taken att he Okayama Astrophysical Observatory and Gunma Astronomical Observatory.
This work was also supported by Research Fellowships for the Promotion of Science for Young Scientists (YM), the Grants-in-Aid for the Global COE Program ``The Next Generation of Physics, Spun from Universality and Emergence" from the Ministry of Education, Culture, Sports, Science and Technology (MEXT) of Japan, and by Grants-in-Aid from MEXT (No. 21740148; SH).


\bsp

\label{lastpage}

\end{document}